\documentclass[preprintnumbers,showpacs, showkeys, 
twocolumn,
secnumarabic,amssymb,amsmath,amsfonts,aps, pra]{revtex4}

\usepackage{graphicx}
\usepackage{mathcomp}

\usepackage{amsmath}
\begin{document}

\newcommand{\non}{\nonumber}
\newcommand{\be}{\begin{equation}}
\newcommand{\ee}{\end{equation}}
\newcommand{\bq}{\begin{eqnarray}}
\newcommand{\eq}{\end{eqnarray}}
\newcommand{\lps}{\langle}
\newcommand{\rps}{\rangle}
\newcommand{\mhn}{\mathfrak{m}^{_{\tiny\textsc{N}}} (V_0)}
\newcommand{\mvnu}{\mathfrak{m}^{_{\tiny\textsc{N+1}}} (V_1)}
\newcommand{\mvhu}{\mathfrak{m}^{_{\tiny\textsc{N+1}}} (V_0,V_1)}
\newcommand{\boldnabla}{\mbox{\boldmath$\nabla$}}
\newcommand{\mvkh}{\mathfrak{m}^{_{\tiny\textsc{N+K}}} ({\bf V})}

\title{Geometric 
Microcanonical Thermodynamics for Systems with First Integrals}

\date{\today}

\author{Roberto Franzosi}
\affiliation{CNR - Istituto dei Sistemi Complessi, via Madonna del Piano 10, 
I-50019 Sesto Fiorentino, Italy}

\begin{abstract}
In the general case of a many-body Hamiltonian system, described by an autonomous
Hamiltonian $H$, and with $K\geq 0$ independent conserved quantities, we derive
the microcanonical thermodynamics. By a simple approach, based on the differential
geometry, we derive  the microcanonical entropy and the derivatives of the entropy
with respect to the conserved quantities. In such a way, we show that all the
thermodynamical quantities, as the temperature, the chemical potential or the
specific heat, are measured as a microcanonical average of the appropriate microscopic
dynamical functions that we have explicitly derived.
Our method applies also in the case of non-separable Hamiltonians, where the usual
definition of kinetic temperature, derived by the virial theorem, does not apply. 
\end{abstract}
\maketitle

In spite traditionally, the microcanonical ensemble has played a minor role respect to
the canonical and gran-canonical approach, in reason of the more complicated formulation
of the statistical formulas, nowadays, thanks to the performance provided by the current
computers, the microcanonical simulation is probably the most used tool in physics.
Molecular dynamics is applied for investigations that cover the fields of the
dynamics of nonlinear semi-classical systems, the living matter issues (protein
folding, DNA-helix), the dynamics of nanosystems and so on.
The ergodicity in fact, whenever holds true, allows to measure dynamic observables as
temporal averages of appropriate functionals along the dynamical evolution, instead
of perform integrals (often unmanageable) on the phase space. 
In this contest have been written very famous papers \cite{LPV-PHT},
that date back to the early applications of molecular-dynamics in physics.
However, their application is restricted to systems described by a separable
Hamiltonian, i.e. given by a standard term of kinetic energy plus an interaction
potential term.
In a series of more recent papers \cite{rugh,rugh1,rugh2}, Rugh has developed a
microthermodynamic formalism to measure thermodynamical functions in the microcanonical
ensemble. The Rugh's formalism applies to non-separable Hamiltonian systems, and to the
case of  Hamiltonians that depend on parameters.

In the present paper, we illustrate a simple geometric approach to measure thermodynamic
observables within the microcanonical ensemble, in systems described also by nonstandard
Hamitonians \cite{note1} and with an arbitrary number of first integrals.
We consider a classical many-particle system described by a
Hamiltonian $H(x^{_1},\ldots,x^{_{N+K}})$ which, in addition to the total energy $H$,
has $K \geq 0$ independent conserved quantities $V_j (x^{_1},\ldots,x^{_{N+K}})=v_j$
such that $\{H,V_j \} = 0$ for $j=1,\dots,K$. All the first integrals are assumed
to be in involution.
\textit{We derive in detail the expressions for the microcanonical  entropy (i.e.
the expression of the microcanonical invariant measure), and we give the general
formulas that allow to measure all the thermodynamic function as the temperature,
the chemical potential, the specific heat, the pressure, and so on.}
More precisely, we show that entropy and the other thermodynamic functions are
derived by multidimensional integrals performed over the sub-manifolds given by the
intersection of the constant energy hyper-surfaces $H(x^{_1},\ldots,x^{_{N+K}})= v_0$
with those defined by $V_j(x^{_1},\ldots,x^{_{N+K}})=v_j$ for $j=1,\dots,K$.
In particular, we show that all order derivatives of the entropy with respect to
the $v_j$ for $j=0,\dots,K$, from which the microcanonical observables depend on,
can be calculated as time averages along the time evolution of appropriate functions,
whenever  the hypothesis of ergodicity holds true.
In order to simplify the notations, let us introduce the following definitions:
with ${\bf x} := (x^{_1},\ldots,x^{_{N+K}})$ we indicate a generic point of the
system phase-space, with $V_0 ({\bf x}) := H({\bf x})$ the ``first'' of the
conserved quantities, and with $\Lambda^{j}_{v_{j}}=\{ {\bf x} \in \mathbb{R}^{_{N+K}}
\mid V_{j}({\bf x})=v_{j}\}$ for $j=0,\dots,K$ the hyper-surfaces associated to
the $K+1$ conserved independent quantities.
We shall assume the level sets of ${V_j}$ to be non-singular hyper-surfaces.
Even if the hyper-surfaces $\Lambda^{j}_{v}$, in general, constitute a singular
foliation, since for some values of $v_j$ an hyper-surface is not a differential
manifold, for generic values of $v_j$ this is not an issue.
Furthermore, we will indicate with
${\bf V}=(V_0,\dots,V_K)$ a vector whose components are the conserved quantities,
and with ${\bf v}=(v_0,\dots,v_K)$ the vector of the corresponding values, for a generic
configuration.
In addition, we will indicate with $\mathfrak{M}_{K}({\bf v}) \subset \mathbb{R}^{_{N+K}}$
the set given by the intersection of the hyper-surfaces $\Lambda^{j}_{v_{j}}$, for
$j=0,\dots,K$, that is $\mathfrak{M}_{K} ({\bf v}) = \bigcap^K_{j=0} \Lambda^{j}_{v_{j}}$.
We shall give the proof of statements reported above, in three steps, according to the
choice $K=0,1$ and for an arbitrary value of $K$.

{\it Case $K=0$.}$-$Let us begin by summarizing the results of Ref. \cite{khinchin,rugh}.
Let ${H}(\mathbf{x})$ be a classical Hamiltonian
describing an autonomous many-body system whose coordinates and
canonical momenta are represented as $N$-component vectors
$\mathbf{x} \in \mathbb{R}^{_{N}}$. Let us suppose
that this system has just one conserved quantity, that is the total
energy $H$.
The microcanonical description of equilibrium thermodynamics for this system,
is given in terms of the microcanonical entropy $S(v_0)$.
Among the equivalent expressions allowed for $S(v_0)$ one can consider the surface
entropy \cite{khinchin} $S(v_0) = \ln\! \int\!\! d^{N}\mathbf{x} 
\delta({V_0}(\mathbf{x})\! -\!v_0)$, in \cite{khinchin,rugh} it is shown the following identity
\begin{equation}
S(v_0) =
\ln\! \int_{\Lambda^0_{v_{0}}}\!\!
\dfrac{    {\mhn}}{\Vert \nabla {V_0} \Vert} \, . \nonumber
\end{equation}
This expression has a precise geometrical interpretation. In fact, 
$
\mhn = \sum_{j=1}^{N} (-1)^{j-1} n^0_j dx_1 \cdots \widehat{dx_j} \cdots dx_{N}
$
is the metric induced from
$\mathbb{R}^{_{N}}$ on the hyper-surface $\Lambda^0_{v_{0}}$,
where $n^0_j ={\partial_j {V_0}}/{\Vert \nabla {V_0} \Vert}$ are the components of
the unitary vector orthogonal to the hyper-surface, and
$1/\Vert \nabla {V_0} \Vert$ is the  microcanonical measure.
The symbol $\widehat{dx_{j}}$, means that $dx_{j}$ has been
lifted from the formula. The microcanonical thermodynamics
is obtained by calculating the derivatives of  $S(v_0)$. This fact allows a geometric
interpretation of all thermodynamic quantity. E.g. the inverse temperature is given
by the definition $1/T(v_0) = \partial S(v_0)/\partial v_0$. This quantity can be rewritten
in a geometric form by using the Federer-Laurence derivation formula 
\cite{federer,laurence,TH1,TH2,PettiniBook}
\begin{equation}
\dfrac{\partial^k}{\partial v_0^k}  \int_{\Lambda^0_{v_{0}}} \mhn  \ \psi
= \int_{\Lambda^0_{v_{0}}} \mhn \  A^k\left(V_0, \psi\right)
\label{lfderiv0}
\end{equation}
where, by setting $n^0 = { \nabla V_0}/{\|\nabla V_0 \|}$, it is
\begin{equation}
A(V_0,\psi) =1/{\| \nabla V_0 \|} {\nabla}
 \left(n^0 \psi \right) \, .
\label{ak0}
\end{equation}
In fact, the inverse microcanonical temperature results (see \cite{rugh} for details)
\begin{equation}
\frac{1}{T(v_0) } = \frac{\int_{\Lambda^0_{v_{0}}} \!\! 
{\mhn}/{\Vert \nabla {V_0} \Vert} 
\nabla
 \left({ n^0}/{\|\nabla V_0 \|} \right)}{\int_{\Lambda^0_{v_{0}}}\!\!
{\mhn}/{\Vert \nabla {V_0} \Vert}} \, , \nonumber
\end{equation}
that is, the microcanonical average of $\nabla
 \left({ n^0}/{\|\nabla V_0 \|} \right)$ on the energy hyper-surface
$\Lambda^0_{v_{0}}$.

{\it Case $K=1$.}$-$Let us consider a system described by the Hamiltonian
$H(x^{_1},\ldots,x^{_{N+1}})$ which, has in addition to the total energy $H$, one independent
conserved quantity $V_1 (x^{_1},\ldots,x^{_{N+1}})=v_1$ such that $\{H,V_1 \} = 0$.
Consistently with the previous case, the microcanonical entropy is given
by  $S({\bf v}) = \ln\! \int\!\! d^{N+1}\mathbf{x} 
\delta({\bf V}(\mathbf{x})\! -\!{\bf v})$, where $\delta({\bf V}(\mathbf{x})\! -\!{\bf v}) = 
\prod_{j=0,1} \delta(V_{j}(\mathbf{x})\! -\!v_{j})$. The second
delta function has been added in order to take in account of the second first-integral of
motion. In \cite{FranzJSP11} it has been shown that 
\begin{equation}
S({\bf v}) =
\ln\! \int_{\mathfrak{M}_{1}}\!\!
\dfrac{\mvhu}{\Vert \Pi_{1} \Vert} \, , \nonumber
\end{equation}
where $\mathfrak{M}_{1}$ is the intersection set of the hyper-surfaces $\Lambda^{0}_{v_{0}}$
and $\Lambda^{1}_{v_{1}}$. The measure $\mvhu$ is that one induced by
$\mathbb{R}^{_{N+1}}$ on $\mathfrak{M}_{1}$, that is
\begin{equation}
\mvhu =\!\!\!\!\! \sum_{{\mu, \nu =1} \choose {\mu < \nu}}^{N+1}\!\! (-1)^{^{^{\!\!\mu-\nu+1}}} 
\!\! \!\!\!\! \!\!\!\! F_{\mu\nu} dx_1 \dots \widehat{dx_\mu}\dots \widehat{dx_\nu} 
\dots  dx_{N+1} \nonumber
\end{equation}
where $F_{\mu\nu} = (e^1_{\mu} e^0_{\nu} - e^0_{\mu} e^1_{\nu} )$ depends on the unitary
vectors of a basis $\{e^0,e^1\}$ are derived by the normalized gradient vectors
$n^j = \nabla V_{j}/\Vert \nabla V_{j} \Vert$ for $j=0,1$, by means of a Gram-Schmidt
orthonormalization process, starting from the vector $e^1=n^1$. Thus, 
$e^0 = [n^{0}-(n^{0} \cdot n^{1}) n^{1}]/[1- (n^{0} \cdot n^{1})^2]^{1/2}$. Furthermore
$\Pi_{1} = { \boldnabla V_0} \wedge {\boldnabla V_{1}}$, where
$\boldnabla V_j = \sum_\mu \partial_\mu V_j dx^\mu$ for $j=0,1$. 
In order to derive the temperature in the microcanonical ensemble, according to the definition 
$T({\bf v}) = \left(\partial S(v_{0},v_{1})/\partial v_{0}\right)^{-1}$, we shall use the
following generalization of the Federer-Laurence derivation formula
(\ref{lfderiv0})-(\ref{ak0}), whose proof is given below.
The generalized derivation formula results \cite{note2}
\begin{equation}
\dfrac{\partial^k}{\partial v_{0}^k} \!\! \int_{\mathfrak{M}_{1}}\!\!\!\! \!\! \mvhu  \ \psi
\!= \!\!\int_{\mathfrak{M}_{1}}\!\!\!\! \mvhu \  A^k_{1}\left(V_0,V_1, \psi\right)
\label{lfderiv1}
\end{equation}
where
\begin{equation}
A_{1}(V_0,V_1,\psi) = \frac{1}{\nabla V_0 \cdot e^{0}}
 \lbrack
\nabla \left( e^{0} \psi \right) -
\psi e^{{1}} \cdot \left( e^{{1}} \cdot \nabla \right) \left(e^{0} \right)
\rbrack
\, , 
\label{ak1}
\end{equation}
is meant as a function of $V_0$ and $V_1$ also through the dependence from these latter
of the unitary vectors $e^0$ and $e^1$.
Here, and in the following we mean $e^{{1}} \cdot (e^{{1}} \cdot \nabla) (e^{0})
= \sum_{j,k}e^{{1}}_{j} e^{{1}}_{k} \partial_{k} (e^{0}_{j})$.

By the calculation reported above, we can easily derive $\partial
S(E,v_{1})/\partial v_1$ that, when $V_1$ is the total number of atoms,
gives the chemical potential via the definition $\mu/T = - \partial
S(E,v_{1})/\partial v_1$.
Indeed, by exchanging $V_0$ with $V_1$ and $v_0$ with $v_1$,  it is easy derive from 
Eq. (\ref{lfderiv1}) the following formula
\begin{equation}
\dfrac{\partial^k}{\partial v_1^k}  \int_{\mathfrak{M}_{1}} \mvhu  \ \psi
= \int_{\mathfrak{M}_{1}} \mvhu \  B^k_{1}\left( V_0,V_1, \psi\right) \nonumber
\end{equation}
where
\begin{equation}
B_{1}(V_0,V_1, \psi) = A_{1}(V_1,V_0, \psi)
\, .
\label{bk1}
\end{equation}
Obviously
\begin{eqnarray}
\dfrac{\partial^{k+q}}{\partial v_1^k \partial v_0^q}  \int_{\mathfrak{M}_{1}} \mvhu 
\ \psi =
\nonumber \\
 \int_{\mathfrak{M}_{1}} \mvhu \  B^k_{1}\left(V_0,V_1 ,A^q_{1}\left(V_0,V_1, \psi\right)
\right) \, .
\label{lfderiE}
\end{eqnarray}

{\it Proof of the generalization of the Federer-Laurence derivation formula.}$-$In this
section we shall give the proof of the generalization of the Federer-Laurence
theorem to varieties of co-dimension two. As first step we shall calculate $\partial
(\int_{\mathfrak{M_{1}}}  \psi \mvhu )/\partial v_0$ and, by iteration we will achieve
our aim. In this derivation we make use of the Stokes theorem. 
Let $U$ be a subset of $\Lambda^{1}_{v}$, such that $\partial U =
\mathfrak{M}_{1}(v_0+\Delta v_0, v)
\cup \mathfrak{M}_{1}(v_0, v)$. The Stokes theorem states
\begin{equation}
\int_{\partial U} \omega = \int_{U} d \omega \, .
\label{ST}
\end{equation}
Let us choose $\omega = \psi {\mvhu}$, thus we obtain
\begin{equation}
d\omega= \sum^{N+1}_{\nu,\mu=1} (-1)^{\nu} \partial_{\mu} 
\left( \psi F_{\mu\nu}\right) dx_1 \dots \widehat{dx_\nu} \dots dx_{N+1}
\nonumber
\end{equation}
and, by using the definition of $ F_{\mu\nu}$, it follows
\[
 \int_U d \omega = 
\int_{U}
\left[ 
 \nabla \cdot \left( \psi e^{0} \right) 
- \psi e^{1} \cdot \left( e^{1} \cdot \nabla \right)\left( e^{0} \right) 
\right] 
\mvnu \, .
\]
Now it is easy to verify that 
\[
\mvnu = \sum_\alpha e^{0}_\alpha d x_\alpha \wedge
\mvhu \, ,
\]
therefore, by choosing vectors orthogonal to $\mathfrak{M}_{1}(v_0, v)$ and
$\mathfrak{M}_{1}(v_0+\Delta v_0, v)$ (but tangent to $\Lambda^1_v$) with opposite orientations, we
get from the Stokes theorem
\begin{eqnarray}
\int_{\mathfrak{M}_{1}(v_0+\Delta v_0, v)} \psi {\mvhu} -
\int_{\mathfrak{M}_{1}(v_0, v)} \psi {\mvhu} 
 \nonumber
=
\\ \nonumber
\int^{\Delta v_0}_0 d s
 \int_{\mathfrak{M}_{1}(v_0 + s, v)}
\mvhu  \times ~~~~~~~~~~~~~~~~~~~~~~~~~~~
\\  \nonumber
 \frac{1}{\nabla V_0 \cdot e^{0}}
      \left[
	   \nabla \cdot \left( \psi e^{0} \right) -
	    \psi e^{1} \cdot \left( e^{1} \cdot \nabla \right)\left( e^{0} \right)
      \right]
\, ,
\end{eqnarray}
that, in the limit $\Delta v_0 \to 0$ gives Eq. (\ref{lfderiv1}) for $k=1$, by iteration
the generalization of the Laurence-Federer formula is proved.

{\it General case.}$-$In this section we will derive the microcanonical entropy,
and its derivatives with respect to the conserved quantities, in the case of systems
with $K > 1$ first integrals in addition to the energy. 

Let a system be described by the Hamiltonian
$H(\mathbf{x})$ that depends on $\mathbf{x} = (x^{_1},\ldots,x^{_{N+K}})$. Let us suppose
the system to have, in addition to the total energy $H$, $K$
independent conserved quantity $V_j (\mathbf{x})=v_j$ such that $\{H,V_j \} = 0$ for
$j=1,\dots,K$.
The microcanonical entropy is given
by  $S({\bf v}) = \ln\! \int\!\! d^{N+K}\mathbf{x} 
 \delta(V(\mathbf{x})\! -\!v)$,
where $ \delta(V(\mathbf{x})\! -\!v) = {\prod^{K}_{j=0}} 
\delta(V_{j}(\mathbf{x})\! -\!v_{j})$. In this case 
\begin{equation}
S({\bf v}) =
\ln\! \int_{\mathfrak{M}_{K}}\!\!
\dfrac{\mvkh}{\Vert \Pi_{K} \Vert} \, , 
\nonumber
\end{equation}
where $\mathfrak{M}_{K}$ is the intersection set of the hyper-surfaces
$\Lambda^{j}_{v_{j}}$ for $j=0,\dots,K$, ${\bf V}=(V_0,\dots,V_K)$ and
${\bf v}=(v_0,\dots,v_K)$.
The measure $\mvkh$ is that one induced
by $\mathbb{R}^{_{N+K}}$ on $\mathfrak{M}_{K}$, and it is defined as follows.
Let $n^{j} = \nabla V_j / \Vert \nabla V_j \Vert $ be the unitary vectors 
orthogonal to the hyper-surfaces $\Lambda^j_{v_j}$, for $j=0,\dots,K$.
By the Gram-Schmidt orthonormalization process, starting from the set of $K+1$
independent unitary vectors $\{ n^0, \dots, n^{K} \}$ we can obtain the following
orthonormalized basis $\{ e^0,\dots, e^{{K}} \}$, where $e^{{K}}
= n^{{K}}$. Therefore,
the measure  $\mvkh$ results
\begin{equation}
 \mvkh = * ({\bf e}^{{K}} \wedge
\cdots \wedge {\bf e}^{{0}}) \, ,
\nonumber
\end{equation}
where ${\bf e}^{j} = \sum_\alpha { e}^{j}_\alpha d x^\alpha $, for $j=0,\dots,K$, and $*$
is the (Hodge) star operator.
The microcanonical measure is the norm of
\begin{equation}
 \Pi_K = { \boldnabla V_0} \wedge { \boldnabla V_1} \wedge
\cdots \wedge { \boldnabla V_K} \, ,
\label{Pik}
\end{equation}
where $\boldnabla V_j = \sum_\mu \partial_\mu V_j dx^\mu$, for $j=0,\dots,K$. 

For convenience let us introduce the following notation:
$({\bf \bar v}) = (v_1, \dots,v_K)$ and $({\bf \bar V}) = (V_1, \dots,V_K)$.
Let us calculate the derivative of entropy with respect to $v_0$, the derivates with
respect to the other variables are obtained easily by cyclic permutation of the indices.
Also here we resort to the Stokes theorem (\ref{ST}) where, in this case, $U$ is a
sub-set of the intersection set of the hyper-surfaces $\Lambda^j_{v_j}$ for $j=1,\dots,K$,
in such a way that $\partial U =  \mathfrak{M}_{K}(v_0+\Delta v_0, {\bf \bar v})
\cup \mathfrak{M}_{K}(v_0, {\bf \bar v})$. Furthermore, we choose $\omega = 
\psi \mvkh$, thus the following facts hold true
\[
\mathfrak{m}^{_{\tiny\textsc{N+K}}} ( {\bf \bar V}) = 
\sum_\alpha e^{0}_\alpha d x_\alpha \wedge
\mvkh \, ,
\]
where $\mathfrak{m}^{_{\tiny\textsc{N+K}}} ( {\bf \bar V}) =
* ({\bf e}^{{K}} \wedge \cdots \wedge {\bf e}^{{1}})$
\begin{eqnarray}
\int_{\mathfrak{M}_{K}(v_0+\Delta v_0,  {\bf \bar v})} \psi {\mvkh} -
\int_{\mathfrak{M}_{K}(v_0,  {\bf \bar v})} \psi {\mvkh} 
 \nonumber
= \\ \nonumber
\int^{\Delta v_0}_0 d s
 \int_{\mathfrak{M}_{1}(v_0 + s,  {\bf \bar v})}
\mvkh  \times ~~~~~~~~~~~~~~~~~~~~~~~~~~~
\\  \nonumber
 \frac{1}{\nabla V_0 \cdot e^{0}}
      \left(
	    d \omega , *
	    ({\bf e}^{{K}} \wedge
\cdots \wedge {\bf e}^{{1}})
      \right)
\, ,
\end{eqnarray}
where $(,)$ is the inner product.
Thus we have
\begin{equation}
\dfrac{\partial^\ell}{\partial v_{0}^\ell}  \int_{\mathfrak{M}_{K}} \mvkh  \ \psi
= \int_{\mathfrak{M}_{K}} \mvkh \  A^\ell_{K}\left({\bf V}, \psi\right)
\label{lfderivk}
\end{equation}
where
\begin{equation}
A_{K}({\bf V},\psi) =
 \frac{1}{\nabla V_0 \cdot e^{0}}
      \left(
	    d \omega , *
	    ({\bf e}^{{K}} \wedge
\cdots \wedge {\bf e}^{{1}})
      \right)
\, .
\label{akk}
\end{equation}

{\it Application to the discrete nonlinear Schr\"odinger equation.}$-$With reference to the
standard canonical coordinates, the Hamiltonian of the discrete nonlinear Schr\"oringer
equation (DNSE) writes
\[
H_\alpha(p,q) = \Lambda_\alpha \sum_{j}  (p_j^2+q_j^2)^2 - \tau_\alpha
 \sum_{j}(p_jp_{j_{+1}}+q_jq_{j_{+1}}) \, ,
\]
where $\Lambda_\alpha$ and $\tau_\alpha$  measure the magnitude of the repulsive on-site
interaction and of the hopping intensity, respectively.
Here, the index $j$ numbers the sites from $1$ to $M$ and periodic boundary
conditions are assumed. This Hamiltonian describes the dynamics of a system
of a Bose-Einstein condensate (BEC) of repulsive-atoms, in an optical lattice
in the superfluid regime. The model possesses two conserved quantities: the energy
$V_0=H_\alpha$ and the number of particles $V_1 = \sum_j (q_j^2 + p_j^2)/2$.
Thus, it corresponds to the case $K=1$ of the present paper. Under the
hypothesis that the ergodicity holds true, for this system the temperature, as a
function of the total energy $E$ and number of atoms $N$, can
be measured as a temporal average along the dynamical evolution. By using
Eq. (\ref{ak1}) we get
\[
 \frac{1}{T(E,N)} = \lim_{s\to \infty} 
 \frac{1}{s} \int_0^s d s^\prime [\Phi(x(s^\prime))] \, ,
\]
where
\begin{multline}
\Phi(x) \! = \! 
		\dfrac{\Vert \Pi_1 \Vert}{ \bigtriangledown H \cdot e^{0} } 
		\left[ 
		 \bigtriangledown \left( 
\frac{e^{0}}{\Vert \Pi_1 \Vert} \right)
- \frac{(e^{1} \cdot \bigtriangledown) \left( e^{0} \right) }{ \Vert \Pi_1 \Vert} 
\cdot
e^{1}
		\right] \, ,
\nonumber
\end{multline}
is written in terms of $e^1 = \nabla V_1/ \Vert \nabla V_1 \Vert$ and
\[
e^0=
{ [n^0 -(n^0 \cdot e^1) e^1]}/{
[1 - (n^0 \cdot e^1)^2]^{1/2}
}
 \, ,
\] 
where $n^0 = \nabla H_\alpha/ \Vert \nabla H_\alpha \Vert$ . Furthermore it results
\[
\Vert \Pi_1 \Vert = \left[ \sum^{_{(N+1)/2}}_{\genfrac{}{}{0pt}{}{\mu,\nu=1}{\mu<\nu}}
\left(
\frac{\partial H_\alpha}{\partial x_\mu} \frac{\partial V_1}{\partial x_\nu} -
\frac{\partial H_\alpha}{\partial x_\nu} \frac{\partial V_1}{\partial x_\mu}
  \right)^2 \right]^{1/2}  \, ,
 \]
with and $x_{2\mu -1}=q_\mu$, $x_{2\mu}=p_\mu$ are the $M=(N+1)/2$ lattice coordinates.
By exchanging $H_\alpha$ with $V_1$, and vice versa, in these equations we can measure
the chemical potential.
In Ref. \cite{Iubini12}, these formulas have been checked and used to explore the
thermodynamics of the DNSE where it has been observed the existence of negative
temperature states. Analogously, in Ref. \cite{Davis} the microcanonical approach
proposed by Rugh, has been applied in the contest of Bose-Einstein condensates,
studying the projected Gross-Pitaevskii equation.

{\it Application to the BEC mixtures.}$-$The dynamics of a mixture of
two bosonic species, condensed in an optical lattice and in the superfluid regime
is described by two coupled DNSEs and the Hamiltonian writes
\[
 H(p,q;\tilde p, \tilde q) = H_1(p,q) + H_2(\tilde p,\tilde q) + 
\beta \sum_j (p^2_j + q^2_j)(\tilde{p}^2_j + \tilde{q}^2_j) \, ,
\]
where $\beta$ takes into account the interaction between the two species, and by assuming
periodic boundary conditions the index $j$ runs from $1$ to $M$. In this case there are
three conserved quantities the total energy $V_0 = H$, and the total number of atoms of
each species: $V_1 = \sum_j (q_j^2 + p_j^2)/2$ and
$V_2 = \sum_j (\tilde{q}_j^2 + \tilde{p}_j^2)/2$. Thus, we have to use the formulas of
the case $K=2$ and, from Eq. (\ref{akk}) the inverse temperature is measured as the
microcanonical average of the function
\begin{eqnarray}
\Phi_2 (x) = 
\frac{\Vert \Pi_2 \Vert}{\nabla H \cdot e^{0}} 
\sum_{\genfrac{}{}{0pt}{}{\mu,\sigma }
{\mu^\prime,\sigma^\prime }}
{\cal A}_{\mu\sigma} e^2_{\mu^\prime} e^1_{\sigma^\prime}
\left( 
\delta_{\mu \mu^\prime} \delta_{\sigma \sigma^\prime} -
\delta_{\mu \sigma^\prime} \delta_{\sigma \mu^\prime}
\right) 
\, , \nonumber
\end{eqnarray}
where $\delta$ is the Kronecker's delta and 
\[
 {\cal A}_{\mu \sigma} = \sum^{N+K}_{\rho = 1}
 \sum^2_{i,j,k=0} \!\!\!\! \varepsilon_{ikj} \left[ 
\partial_{\rho} (\psi) (e^i_{\rho} e^j_\mu e^k_\sigma) +
\psi \partial_{\rho} (e^i_{\rho} e^j_\mu e^k_\sigma)
\right] \, .
\]
Here, $\varepsilon$ is the completly antisymmetric tensor, the unitary vectors are
$e^j = \nabla V_j/ \Vert \nabla V_j \Vert$ for $j=1,2$,
and $e^0 = \eta / \Vert \eta \Vert$, where
$$
\eta = n^0 - (n^0 \cdot e^1) e^1 - (n^0 \cdot  e^2) e^2 \, ,
$$
and $n^0 = \nabla H/ \Vert \nabla H \Vert$.
Finally, from Eq. (\ref{Pik}) we have
\[
\Vert \Pi_2 \Vert = \left[ \sum^{_{(N+2)/2}}_{\genfrac{}{}{0pt}{}{\mu_i,\mu_j ,\mu_k =1}
{\mu_i<\mu_j<\mu_k}}
\left(\sum^3_{i,j,k=1} \varepsilon_{ijk}
\frac{\partial H}{\partial x_{\mu_i}}
\frac{\partial V_1}{\partial x_{\mu_j}}
\frac{\partial V_2}{\partial x_{\mu_k}}
  \right)^2 \right]^{1/2}  \, ,
 \]
with $x_{2\mu-1}=q_\mu$, $x_{2\mu}=p_\mu$, for $\mu = 1,\dots,(N+2)/4$, and 
$x_{2\mu-1}=\tilde{q}_{\mu-M}$, $x_{2\mu}=\tilde{p}_{\mu-M}$, for
$\mu = (N+2)/4+1,\dots,(N+2)/2$. In this case the number of lattice sites is
$M= (N+2)/4$.
These equations are presently used to investigate the thermodynamics of two-component
bosons mixtures in Ref. \cite{FranzFuture}.

{\it Final remarks.}$-$In this paper we have proposed a geometric approach to
measure thermodynamic observables within the microcanonical ensemble in classical
Hamiltonian systems with $K \geq 0$ independent conserved quantities. The method
that we have shown is applicable also to systems with nonstandard Hamiltonians and
with an arbitrary number of conserved quantities. As an example we have derived the
formula of the temperature for a model that describes the superfluid dynamics of
two-components bosons on an optical lattice.

{\it Acknowledgment.}$-$I thank Prof. Giovanni Lombardi for the helpful 
discussions at I.P.S.I.A. C. Cennini.

\hfill






\bibliographystyle{elsarticle-num}
\bibliography{<your-bib-database>}

\begin{thebibliography}{99}
\bibitem{note1} With nonstandard we mean Hamiltonians that are not quadratic in the
momentum variable.
\bibitem{note2} These formulas have been derived for the first time in Ref. 
\cite{FranzJSP11}, where there are reported also some applications.
\end{thebibliography}




\end{document}